\documentclass{llncs}
\usepackage{makeidx}  
\usepackage{graphicx}
\usepackage[ampersand]{easylist}
\usepackage{multirow}
\begin{document}
\frontmatter          
\pagestyle{headings}  
\addtocmark{Hamiltonian Mechanics} 
\title{Audio-replay attack detection countermeasures}
\titlerunning{Audio-replay attack detection countermeasures}  
%
\author{Galina Lavrentyeva\inst{1} \and Sergey Novoselov\inst{1} \and Egor Malykh\inst{1} \and Alexander Kozlov\inst{1} \and Oleg Kudashev\inst{1, 2} \and Vadim Shchemelinin\inst{1}}
%
%
%
\institute{ITMO University, St.Petersburg, Russia\\
\and
STC-innovations Ltd., St.Petersburg, Russia\\
\email{\{lavrentyeva, novoselov, malykh, kozlov-a, kudashev, shchemelinin\}@speechpro.com}}

\maketitle              

\begin{abstract}
This paper presents the Speech Technology Center (STC) replay attack detection systems proposed for Automatic Speaker Verification Spoofing and Countermeasures Challenge 2017. In this study we focused on comparison of different spoofing detection approaches. These were GMM based methods, high level features extraction with simple classifier and deep learning frameworks. Experiments performed on the development and evaluation parts of the challenge dataset demonstrated stable efficiency of deep learning approaches in case of changing acoustic conditions. At the same time SVM classifier with high level features provided a substantial input in the efficiency of the resulting STC systems according to the fusion systems results.

\end{abstract}
\section{Introduction}
In recent years, due to increasing security concerns in all aspects of our daily lives, the need for convenient and non-intrusive authentication methods has grown. 
Automatic speaker verification (ASV) offers a low-cost and reliable solution for identification problem when voice services are provided. It is already used in social security entitlement, immigration control and election management. Speaker recognition systems are widely used in customer identification during call to a call center, Internet-banking systems and other fields of e-commerce. However, despite the fact that it has reached the point of mass market adoption this technology is acknowledged to be vulnerable to spoofing attacks \cite{soton370091}.

According to the \cite{FaundezZanuy} ASV spoofing attacks can be classified into direct and indirect attacks acсording to the stage they are applied to. Indirect ones require access to the system and attack inner modules (feature extraction module, voice models, classification results). In opposite, direct attacks use the recording stage or transmission level and is focused on the substitution of the input data. Since speaker verification is mostly used in automatic systems without face-to-face contact, direct attacks are more likely to be used by criminals due to implementation simplicity. The most well-known spoofing attacks  are impersonation, replay attack (RA), voice conversion (VC) attack and text-to-speech (TTS) attack \cite{villalba2010speaker}, \cite{Wu2015130}. 

Unlike other spoofing types, impersonation does not leave any traces of the recording and playback devices or signal processing, as it is genuine speech of a non-target speaker. It can be detected by reliable speaker verification system \cite{imperson}. TTS and VC detection methods were the topic of Automatic Speaker Verification Spoofing (ASVspoof) 2015 Challenge \cite{wu2015asvspoof}. Results presented during that challenge confirmed the great potential in detection of VC and TTS. Compared to these spoofing types RA is much more simpler as it's realisation does not require specific audio signal processing knowledges. In RA fraudster usually uses pre-recorded speech samples of the target speaker that can be easily prepared via low-cost recording devices or smartphones. Due to this RA is the most available and therefore critical spoofing technique.

For today there is a small number of studies addressed to the RA detection. The most part of solutions presented for text-dependent ASV are based on the comparison of the test utterance with the stored utterance recorded during the registration. Vulnerability of text-independent ASV to RA was considered in \cite{villalba2010speaker} and \cite{villalba2011preventing}. These papers show the serious increase in false acceptance rate of ASV system in case of RA presence. RA detection methods for text-independent case are mostly based on additional noise detection, specific for pre-defined acoustic conditions.

The second edition of ASVspoof Challenge conducted in 2017 was aimed to promote the development of RA countermeasures reliable to both known and unknown conditions \cite{ASVspoof2017} which can vary greatly. ASVspoof 2017 was focused on a standalone RA detection task for text-dependent case considered without ASV system and any pre-recorded enroll data.

In this paper we described Speech Technology Center (STC) RA detection systems proposed for ASVspoof 2017. Here we investigated and compared different approaches for spoofing detection. These were Gaussian Mixture Model (GMM) based systems, systems based on high level features with simple classifier and deep learning approaches.

%
\section{Automatic Speaker Verification Spoofing Detection Challenge 2017}
ASVspoof Challenge was organized in order to assess the potential to detect RA "in the wild", specifically in varying acoustic conditions. For this purpose the spoofing database was collected using text-dependent RedDots data \cite{RedDots}. RedDots corpus was replayed and recaptured in heterogeneous acoustic environment (open lab space, balcony, etc.). For spoofing trials 15 different playback and 16 recording devices were used, including smartphones and high-quality speakers. The original RedDots records were used as genuine speech trials. This dataset was divided into training, development and evaluation parts. The evaluation part contained no information about spoofing trials, devices and recording conditions. Spoofing trials from evaluation part were prepared with the use of devices that were not used in the recording process of the training or development data. In this way they presented previously unforeseen spoofing attacks.
\section{GMM approach}
Authors of \cite{CQT} insist that spoofing detection methods should be more focused on the feature engineering rather than on models complication. They proposed the system based on constant Q transform cepstral coefficients (CQCC) and simple GMM. This approach showed impressive results on the ASVspoof 2015 dataset and achieved the 72\% improvement over the best system participated in the challenge.
That is why the reference implementation of this system was provided as a Baseline system by organisers of ASVspoof 2017.
\subsection{Baseline}
\begin{figure}[]
  \centering
  \begin{minipage}[b]{0.52\textwidth}
    \includegraphics[width=\textwidth]{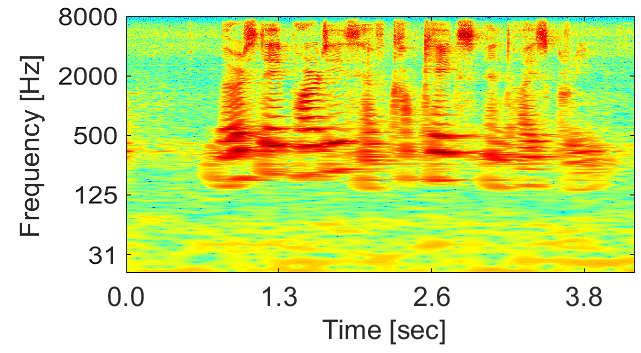}
    \caption{Log power magnitude CQT spectrum}
    \label{fig:spectr}
  \end{minipage}
  \hfill
  \begin{minipage}[b]{0.43\textwidth}
    \includegraphics[width=\textwidth]{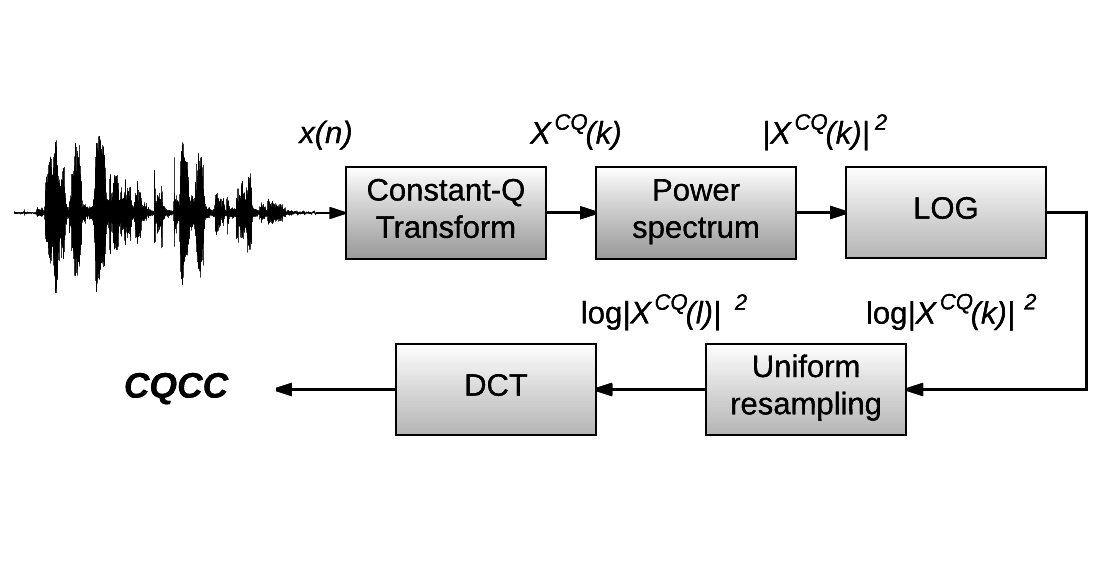}
    \caption{CQCC features extraction}
    \label{fig:CQCC}
  \end{minipage}
\end{figure}

Constant Q transform (CQT) is widely used in music signal processing. By using geometrically spaced frequency bins it overcomes the lack of frequency resolution at lower frequencies and time resolution at higher frequencies that can be produced by Fourier transform with regular space frequency bins. Figure \ref{fig:spectr} demonstrates the example of CQT spectrum. CQCC estimated according to the scheme in Figure \ref{fig:CQCC} were used as input features in the Baseline system.

In the Back-End the Baseline system used standard 2-class GMM classifier. 512-component models were trained with an expectation-maximisation (EM) algorithm with random initialisation for genuine and spoofed speech, respectively. For each utterance the log-likelihood score was obtained from GMM models and the final score was computed as the log-likelihood ratio: $\Lambda(X) = \log L(X \vert \theta_g) - \log L(X \vert \theta_s)$, where $X$ is a sequence of test utterance feature vectors, $L$ denotes the likelihood function, and $\theta_g$ and $\theta_s$ represent the GMM for genuine and spoofed speech.
\subsection{Baseline modifications}
The first modifications we tried were normalization techniques for acoustic features. We considered mean and variance normalization on the log power spectrum (mvn) and cepstral coefficients (cmvn) in different combinations.
We have also tried acoustic features that were effective for VC and TTS detection. These were Mel Wavelet Packet Coefficients (MWPC), based on applying the multiresolution wavelet transform and phase-based features CosPhasePC \cite{STC_ASVspoof2015}.

Experiments results obtained on the development part (Table \ref{tab:baseline}) confirm the efficiency of normalization for CQCC based systems. Systems with other Front-End features did not outperform these results.
\begin{table}[]
\centering
\caption{Experimental results for GMM-based systems obtained on the development and evaluation parts of ASVspoof 2017 database (EER \%)}
\begin{tabular}{|c|c|c|c|c|c|c|c|}
\hline
\textbf{Features} & \multicolumn{4}{c|}{\textbf{CQCC}} & \textbf{MWPC} & \textbf{MWPC} & \textbf{CosPhasePC} \\ \hline
\textbf{Normalization} & - & mvn & cmvn & mvn+cmvn & - & cmvn & - \\ \hline
\textbf{ train EER (\%) } & 0.03 & 1.81 & 1.25 & 1.1  & 0.4 & 0.23  & 0.43 \\ \hline
\textbf{ dev EER (\%) } & 10.35 & 8.74 & 11.86 & 9.85 & 8.81 & 19.17 & 24.64 \\ \hline
\end{tabular}
\label{tab:baseline}
\end{table}

\section{I-vector based system}
\subsection{$\textrm{SVM}_\textrm{i-vector}$}
The most efficient spoofing detection systems on ASVspoof 2015 were based on high level features modelling by standard i-vector approach \cite{TVJFA}. Inspired by its success for VC ans TTS spoofing detection we proposed similar approach for RA detection. We experimented with variety of acoustic features from ASVspoof 2015, such as MFCC, CosPhasePC and MWPC features. According to our observations the best system was the system based on the Linear Prediction Cepstral Coefficients (LPCC).

78 LPCC coefficients were obtained using the Hanning window function with a $0.128$ seconds window size and a $0.016$ seconds step for FFT power spectrum estimation \cite{Ellis05-rastamat}.
I-vectors of the Total Variability space were extracted from the whole speaker utterances. Here we used the the 128-component Gaussian mixture model of the described features for diagonal covariance UBM (Universal Background Model), and the dimension of the T-matrix was 200. 
After centering and length-normalization i-vectors were used as an input for a linear kernel SVM classifier. For SVM training the efficient LIBLINEAR \cite{liblinear} library was used.
\subsection{Phrase-dependent system}
During our investigations we suggested that phrase-dependent system can perform a higher accuracy than text-independent system. We compared i-vector based system with the similar systems trained for several different phrases independently. 
Our experiments presented in Table \ref{tab:phrase} showed reduction in spoofing detection in comparison with the common systems trained on the whole training dataset. This effect can be explained by the insufficient size of the training data in a phrase-dependent case which leads to fast overfitting.
\begin{table}[]
\centering
\caption{Experimental results for the i-vector based system obtained on the development and evaluation parts of ASVspoof 2017 database (EER \%)}
\label{tab:phrase}
\begin{tabular}{|l|c|c|c|c|}
\hline
\textbf{UBM} & phrase-dependent & common & common & common \\ \hline
\textbf{T-matrix} & phrase-dependent & phrase-dependent & common & common \\ \hline
\textbf{SVM} & phrase-dependent & phrase-dependent & phrase-dependent & common \\ \hline
\textbf{train EER (\%)} & 0.398 & 0.133 & 0.401 & 1.459 \\ \hline
\textbf{dev EER (\%)} & 11.702 & 12.368 & 11.447 & 9.95 \\ \hline
\end{tabular}
\end{table}

\section{Deep learning approaches}
Deep learning approaches have already achieved remarkable performance in many classification and recognition tasks. The success of CNN in video classification \cite{karpathy2014large}, image classification \cite{bengio2013representation,krizhevsky2012imagenet}, face recognition \cite{taigman2014deepface} prompts to apply such approaches for ASV anti-spoofing tasks. 
The idea of employing CNN for spoofing detection is not new and was used in face spoofing detection \cite{galbally2014biometric}, \cite{yang2014learn}. 

Several experimental results confirm the efficiency of CNN based approaches for synthetic speech detection. For example in \cite{cnnrnn} authors demonstrated high efficiency of deep neural network (DNN), CNN and recurrent neural network (RNN) architectures for VC and TTS spoofing detection on the base of ASVspoof 2015 dataset. They also proposed a stacked CNN+RNN architecture and demonstrated its state-of-the-art performance. It is particularly important to note that CNN architecture showed similar to CNN+RNN results and their fusion outperformed the best individual system twice in terms of detection quality.
In \cite{tempcnn} temporal CNN architectures were proposed for VC and TTS detection. This approach also achieved notable results on the ASVspoof 2015 corpora.

In this section we describe several systems based on CNN frameworks for RA spoofing detection. Such problem can be reduced to the detection of local spectral artifacts presented in the reproduced replay attack that distinguish it from the genuine speech. For this purpose the CNN was used as a robust feature extractor from the input signal representation in a time-frequency domain. 

\subsection{Unified shape time-frequency features}
In our research we chose 2 types of features based on spectrum estimation of the utterance.
To prepare CNN input acoustic features we used the normalized log power magnitude spectrum obtained via:
\begin{itemize}
    \item constant Q transform (CQT) \cite{CQT} 
    \item Fast Fourier Transform (FFT)
    \item Discrete wavelet Transform (DWT), obtained by Daubechies wavelets db4
\end{itemize}

Special attention should be paid here to the fact that CNN input data should have a unified form. We considered two techniques for obtaining a unified time-frequency (T-F) shape of features. First one truncates the spectrum along the time axis with a fixed size. During this procedure short files are extended by repeating their contents if necessary to match the required length. The other technique uses a sliding window approach with a fixed window size.

\subsection{Deep Learning architectures}
It is known from image processing that the choice of convolutional neural network architecture is a critical task and greatly affects learning result.
In this research we investigated several deep neural network architectures, demonstrated the best results in RA detection.
\subsubsection{Inception CNN based system $\mathrm{ICNN}_\mathrm{CQT}^\mathrm{SW}$}
The first proposed neural network architecture was CNN with inception modules (ICNN) (\cite{GoogLeNet}). 
The proposed architecture was the reduced version of GoogLeNet and contained 3 inception modules. Inception module acts as multiple convolution filter inputs, that are processed on the same input in parallel. It also does pooling at the same time. All resulting feature maps are then concatenated before going to the next layer. This allows the model to pick the best convolutions and take advantages of multi-level feature extraction by recovering both local feature via smaller convolutions and high abstracted features with larger convolutions.

The ICNN was applied for high level feature extraction from the log power magnitude CQT spectrum. The last fully-connected (FC) layer with softmax activation function was used to discriminate between genuine and spoofing classes during the training process. And a low-dimensional high-level audio representation was extracted from the penultimate FC layer. 

To obtain unified time-frequency representation of the audio-signal we used sliding window technique with $864 \times 200 \times 1$ window size and 90\% overlapping. For ICNN input we applied mean variance normalization.

Since in this space of high level features genuine and replay spoofing classes are well separated, it was enough to use the simplest one-component models for each class distribution modelling. We used the standard 2-class GMM classifier (1 GMM for genuine speech and 1 for spoofed speech) trained on the training part of the ASVspoof 2017 database with EM algorithm. The score for an input signal was computed as the loglikelihood ratio. In this scenario we extracted high-level features independently for each sliding window and all high-level features corresponding to one utterance were used to estimate GMM likelihood.
However, it should be mentioned that such deep neural approach can be used for End-To-End solution without additional classifier.
\subsubsection{Light CNN}
The second CNN we explored was the reduced version of the LCNN proposed in  \cite{lightcnn} with a smaller number of filters in each layer. LCNN consisted of 5 convolution layers, 4 Network in Network (NIN) layers \cite{lin2013network}, 10 Max-Feature-Map layers, 4 max-pooling layers and 2 fully connected layers \cite{STC_ASVspoof2017}.
The proposed LCNN used Max-Feature-Map activation function that allows to reduce CNN architecture. In contrast to commonly used Rectified Linear Unit function that suppresses a neuron by a threshold (or bias), MFM suppresses a neuron by a competitive relationship. In this way being applied in particular MFM selects more informative features. 
Each convolution layer was a combination of two independent convolutional parts calculated for layer's input. MFM activation function was used then to calculate element-wise maximum of these parts. Max-Pooling layers with kernel of size $2 \times 2$ and stride of size $2 \times 2$ were used for time and frequency dimensions reduction. Described CNN was used to obtain high-level audio features similar to the ICNN based system and simple GMM classifier was used at the evaluation stage.

We proposed a number of systems based on LCNN high-level features extractor with different acoustic features. $\mathrm{LCNN}_\mathrm{CQT}$ used truncated features obtained from the normalized CQT spectrograms with $864 \times 400 \times 1$ size. Additionally we explored FFT based features instead of CQT: $\mathrm{LCNN}_\mathrm{FFT}$ system used truncated features of size $864 \times 400 \times 1$ and $\mathrm{LCNN}_\mathrm{FFT}^\mathrm{SW}$ was based on the sliding window features extraction with $864 \times 200 \times 1$ window and 90\% overlapping along time axis. Alternative system $\mathrm{LCNN}_\mathrm{DWT}^\mathrm{SW}$ was based on DWT implementation with sliding window of $256 \times 200 \times 1$ size and 83.4\% overlapping.
\subsubsection{Stacking CNN and RNN}
We also probed the combined CNN + RNN architecture from \cite{cnnrnn} for RA spoofing detection. In this stacked architecture CNN was used as a feature extractor and RNN modeled the long-term dependencies of a speech sequence. Both CNN and RNN were optimized jointly through the back-propagation algorithm. In this implementation CNN+RNN was used as End-to-End solution.

CNN itself was a reduced version of LCNN. But unlike LCNN from the previous systems max-pooling was applied with the stride $2$ along the frequency axis to compress frequency information and stride $1$ along the time axis to save time dimensionality. 

The RNN part consists of two gated recurrent units \cite{gru} forming the bidirectional gated recurrent unit (BGRU). The first GRU was responsible for the forward pass, while the second GRU performed the backward pass. The last output vectors of both forward and backward passes were taken further to obtain two $16$-dimensional vectors. Such BGRU unit was applied to each channel of CNN's output resulting in $16 \times 2 \times 8$ tensor. Weights were shared between each channel's unit to prevent overfitting.
The flattened output of RNN was used as an input to the fully-connected layers with MFM activations resulting in probability of the utterance being spoofed.

System based on this architecture, $\mathrm{CNN}_\mathrm{FFT}+\mathrm{RNN}$, used truncated features extracted from log magnitude power FFT spectrum. But due to the limited computational resources we reduced the input data dimension to $256 \times 400 \times 1$.

Alternatively we used CNN+RNN architecture for $\mathrm{CNN}_\mathrm{\Delta EEMD}+\mathrm{RNN}$ system based on the ensemble empirical mode decomposition (EEMD) features.These features were obtained by the following algorithm with the use of libEEMD library \cite{libeemd}:
\begin{enumerate}
    \item Let $S_o$ be the FFT spectrogram of the original signal $x(t)$
    \item Get the first empirical mode $c_1(t)$ of the signal $x(t)$ using the EEMD with ensemble size of $50$ and noise strength equals to $0.1 \sqrt{\mathrm{Var}({x(t)})}$ 
    \item Compute $S_r$ as the FFT spectrogram of the signal $c_1(t)$
    \item $S_{\Delta} = \vert S_o - S_r \vert$
\end{enumerate}
\section{Evaluation results and discussion}
\label{sec:Exp}
The experimental results of all described individual systems on development and evaluation parts of ASVspoof 2017 corpus are presented in Table \ref{tab:res}.

\begin{table}[]
  \caption{Evaluation results}
  \label{tab:res}
  \centering
  \begin{tabular}{|l|c|c|}
  \hline
    \multirow{2}{*}{\textbf{Individual system}} & \multicolumn{2}{c|}{\textbf{EER} (\%)}\\
    \cline{2-3} 
     & \textbf{Dev set} & \textbf{Eval set} \\
    \hline
    $\mathrm{Baseline}$ & 10.35 & 30.17 \\\hline
    $\mathrm{Baseline}_\mathrm{MVN+CMVN}$ & 9.85 & 17.4 \\\hline
    $\mathrm{SVM}_\mathrm{i-vec}$ & 9.95 & 12.82 \\\hline
    $\mathrm{ICNN}_\mathrm{CQT}^\mathrm{SW}$ & 5.73 & 15.11\\\hline
    $\mathrm{LCNN}_\mathrm{FFT}$ & \textbf{4.53} & \textbf{7.34} \\\hline
    $\mathrm{LCNN}_\mathrm{FFT}^\mathrm{SW}$ & 5.25 & 11.91 \\\hline
    $\mathrm{LCNN}_\mathrm{CQT}$ & 4.80 & 16.54 \\\hline
    $\mathrm{LCNN}_\mathrm{DWT}$ & 8.71 & 16.41 \\\hline
    $\mathrm{CNN}_\mathrm{FFT} + \mathrm{RNN}$ & 7.51 & 10.68 \\\hline
    $\mathrm{CNN}_\mathrm{\Delta EEMD} + \mathrm{RNN}$ & 9.94 & 18.93 \\\hline
    \textbf{Fusion system}& \textbf{Dev set} & \textbf{Eval set}\\\hline
    Primary: $\mathrm{LCNN}_\mathrm{FFT}$, $\mathrm{SVM}_\mathrm{i-vec}$, $\mathrm{CNN}_\mathrm{FFT}+\mathrm{RNN}$
    & \textbf{3.95} & \textbf{6.73} \\\hline
    Contrastive: $\mathrm{SVM}_\mathrm{i-vec}$, $\mathrm{ICNN}_\mathrm{CQT}^\mathrm{SW}$, $\mathrm{LCNN}_\mathrm{FFT}$, $\mathrm{LCNN}_\mathrm{FFT}^\mathrm{SW}$, & \multirow{2}{*}{2.62} & \multirow{2}{*}{7.56} \\
    $\mathrm{LCNN}_\mathrm{DWT}^\mathrm{SW}$, $\mathrm{CNN}_\mathrm{FFT}+\mathrm{RNN}$, $\mathrm{CNN}_\mathrm{\Delta EEMD}+\mathrm{RNN}$ &  &  \\ \hline
\end{tabular}
\end{table}

The best result for development and evaluation sets was demonstrated by LCNN system with FFT truncated features. Similar system with CQT-based features showed poor stability on the evaluation set. This can be explained by the poor robustness of the CQT features, which is also approved by results of the baseline system.

The sliding window technique demonstrated worse results compared to the truncated approach on the evaluation set. A possible reason for this is that using spectrograms of the whole utterances (in most cases) as CNN input leads to more accurate text-dependent deep model.

Our CNN+RNN combination also performed worse RA detection quality than a single LCNN. We explain performance degradation by the reduced frequency resolution in the spectrum estimation.

Summarizing all results of our individual systems for the final submission of the ASVspoof 2017 Challenge we prepared two solutions that use fusion of individual systems on the score level with Bosaris toolkit \cite{Bosaris}.
Our primary system used fusion of $\mathrm{LCNN}_\mathrm{FFT}$, $\mathrm{SVM}_\mathrm{i-vect}$ and $\mathrm{CNN}_\mathrm{FFT}+\mathrm{RNN}$ systems scores. And our Contrastive system combined $\mathrm{SVM}_\mathrm{i-vec}$, $\mathrm{ICNN}_\mathrm{CQT}^\mathrm{SW}$, $\mathrm{LCNN}_\mathrm{FFT}$, $\mathrm{LCNN}_\mathrm{FFT}^\mathrm{SW}$,  $\mathrm{LCNN}_\mathrm{DWT}^\mathrm{SW}$, $\mathrm{CNN}_\mathrm{FFT}+\mathrm{RNN}$ and $\mathrm{CNN}_\mathrm{\Delta EEMD}+\mathrm{RNN}$ systems.
Comparing two fusion systems on the development and evaluation parts we can see that complicated fusion of 7 systems have less performance than simpler fusion of $\mathrm{LCNN}_\mathrm{FFT}$, $\mathrm{SVM}_\mathrm{i-vec}$, $\mathrm{CNN}_\mathrm{FFT}+\mathrm{RNN}$ which are completely different in architectures, features and presumably detect different artefacts. Despite the noticeable quality reduction for some individual systems on the eval part the difference in 1\% of EER for fusion systems can be explained by the impressive results of $\mathrm{LCNN}_\mathrm{FFT}$. According to Bosaris fusion model, this system has the biggest weight in both fusion solutions.However, we suppose that in case of more common RA detection problem with more various conditions complex system will have better spoofing detection quality.

\section{Conlusions}
\label{sec:conc}

In this paper we explored the applicability of several different approaches for replay attack spoofing detection. We investigated state-of-the-art methods from VC and TTS spoofing detection and deep learning approaches. Our experiments conducted on the ASVspoof 2017 dataset confirmed high efficiency of deep learning frameworks for spoofing detection "in the wild". EER of the best individual CNN system was 7.34\%. At the same time SVM classifier with high level features provides a substantial input into the efficiency of the resulting STC systems according to the fusion systems results. Our primary system based on systems score fusion provided 6.73\% EER on the evaluation set.

%
%
%
\bibliographystyle{ieeetr}
\bibliography{main}
\end{document}